\documentclass[12pt]{article}
\usepackage{amsmath,amssymb,epsfig}
\makeatletter \@addtoreset{equation}{section} \makeatother
\renewcommand{\theequation}{\thesection.\arabic{equation}}
\newcommand{\ba}{\begin{array}}
\newcommand{\ea}{\end{array}}
\newcommand{\beq}{\begin{equation}}
\newcommand{\eeq}{\end{equation}}
\newcommand{\bea}{\begin{eqnarray}}
\newcommand{\eea}{\end{eqnarray}}

\def\bce{\begin{center}}
\def\ece{\end{center}}

\def\nonu{\nonumber}

\def\pa{\partial}
\def\al{\alpha}
\def\be{\beta}

\def\de{\delta}

\def\la{\lambda}

\def\eps6{{\displaystyle \mathop{\epsilon}^{6}}{}}

\def\nab6{{\displaystyle \mathop{\nabla}^{6}}{}}

\def\0{{\sst{(0)}}}
\def\1{{\sst{(1)}}}
\def\2{{\sst{(2)}}}
\def\3{{\sst{(3)}}}
\def\4{{\sst{(4)}}}
\def\5{{\sst{(5)}}}
\def\6{{\sst{(6)}}}
\def\7{{\sst{(7)}}}
\def\8{{\sst{(8)}}}

\def\ba{\begin{array}}
\def\ea{\end{array}}
\def\beq{\begin{equation}}
\def\eeq{\end{equation}}
\def\be{\begin{equation}}
\def\ee{\end{equation}}

\def\Tr{\mathop{\rm Tr}}

\def\la{\lambda}
\def\eps{\epsilon}

\def\ba{\begin{array}}
\def\ea{\end{array}}
\def\beq{\begin{equation}}
\def\eeq{\end{equation}}
\def\be{\begin{equation}}
\def\ee{\end{equation}}

\def\Tr{\mathop{\rm Tr}}

\def\la{\lambda}
\def\eps{\epsilon}

\newcommand{\bean}{\begin{eqnarray*}}
\newcommand{\eean}{\end{eqnarray*}}

\begin{document}
\thispagestyle{empty} \addtocounter{page}{-1}
   \begin{flushright}
\end{flushright}

\vspace*{1.3cm}
  
\centerline{ \Large \bf  The Gauge Dual of  }
\vspace{.3cm} 
\centerline{ \Large \bf  
Gauged ${\cal N}=8$ Supergravity Theory   } 
\vspace*{1.5cm}
\centerline{{\bf Changhyun Ahn }
} 
\vspace*{1.0cm} 
\centerline{\it  
Department of Physics, Kyungpook National University, Taegu
702-701, Korea} 
\vspace*{0.8cm} 
\centerline{\tt ahn@knu.ac.kr
} 
\vskip2cm

\centerline{\bf Abstract}
\vspace*{0.5cm}

The most general $SU(3)$-singlet space of
gauged ${\cal N}=8$ supergravity in four-dimensions is studied recently.
The $SU(3)$-invariant six scalar fields are realized by
six real four-forms.
A family of holographic ${\cal N}=1$ supersymmetric RG flows on
M2-branes in three-dimensions is described. 
This family of flows is driven by three independent mass parameters
from the ${\cal N}=8$ $SO(8)$ theory and is controlled by
two IR fixed points, ${\cal N}=1$ $G_2$-invariant one and 
${\cal N}=2$ $SU(3) \times
U(1)$-invariant one. 
The generic flow with arbitrary mass parameters is
${\cal N}=1$ supersymmetric and reaches to the ${\cal N}=2$ $SU(3) \times U(1)$
fixed point where the three masses become identical. 
A particular ${\cal N}=1$ supersymmetric $SU(3)$-preserving RG flow
from the ${\cal N}=1$ $G_2$-invariant fixed point to the ${\cal N}=2$
$SU(3)\times U(1)$-invariant fixed point is also discussed.   

\baselineskip=18pt
\newpage
\renewcommand{\theequation}
{\arabic{section}\mbox{.}\arabic{equation}}

\section{Introduction}

The
three-dimensional ${\cal N}=6$ $U(N) \times U(N)$ 
Chern-Simons matter theories
with level $k$ 
can be regarded as the low energy limit of $N$ M2-branes at 
${\bf C}^4/{\bf Z}_k$ singularity \cite{ABJM}.
The coupling of this theory may be thought of as $\frac{1}{k}$
and so this is weakly coupled for large $k$. 
For $k=1, 2$, the full ${\cal N}=8$ supersymmetry
is preserved with $SO(8)$ R-symmetry 
and this becomes strongly coupled theory. For $k > 2$,
the supersymmetry is broken to ${\cal N}=6$ and R-symmetry is broken
to $SO(6)$. 

The renormalization group(RG) flow
between the ultraviolet(UV) fixed point and 
the infrared(IR) fixed point of the three-dimensional 
field theory can be described from gauged ${\cal N}=8$ 
supergravity theory in four-dimensions via AdS/CFT 
correspondence \cite{Maldacena,Witten98,GKP}. 
The holographic supersymmetric
RG flow equations connecting ${\cal N}=8$ $SO(8)$-invariant fixed point 
to ${\cal N}=2$ $SU(3) \times U(1)$-invariant fixed point have been found in 
\cite{AP,AW}(See also \cite{NW} for earlier work). 
The other holographic supersymmetric
RG flow equations from ${\cal N}=8$ $SO(8)$-invariant fixed point 
to
${\cal N}=1$ $G_2$-invariant
fixed point also have been studied in 
\cite{AW,AI,AR99}(See also \cite{Warner83,AI02} for previous work on
the critical point and the metric respectively).
The exact solutions to the $M$-theory lift of these supersymmetric
RG flows have been constructed in \cite{CPW,AI} respectively.
There exist three supersymmetric critical points in gauged ${\cal
N}=8$ supergravity theory.

The mass deformed $U(2) \times U(2)$
Chern-Simons matter theory with level $k=1$ 
or $k=2$ preserving global $SU(3) \times U(1)$ symmetry has been studied 
in \cite{Ahn0806n2,BKKS,KKM,KPR} by adding a single mass term for ${\cal N}=2$
superfield.  
The mass deformation for this theory preserving $G_2$
symmetry  has been described in \cite{Ahn0806n1} by adding a single
mass term for ${\cal N}=1$ superfield. 
The nonsupersymmetric 
RG flow equations preserving $SO(7)^{\pm}$ symmetry 
have been discussed in \cite{Ahn0812} by looking at the previous work \cite{AW} closely.  
The holographic
RG flow equations connecting ${\cal N}=1$ $G_2$-invariant fixed point 
to ${\cal N}=2$ $SU(3) \times U(1)$-invariant fixed point have been 
found in \cite{BHPW} by computing and analyzing the mass eigenvalues in gauged
${\cal N}=8$ supergravity. This is the last flow connecting the remaining
two nontrivial supersymmetric critical points. 
Moreover, the ${\cal N}=4$ and ${\cal N}=8$ supersymmetric RG flows have been 
studied in \cite{AW09} by adding the appropriate mass terms.

The gauged ${\cal N}=8$ supergravity in four-dimensions
has a scalar potential which
is a function of 70 scalars, in general \cite{CJ}. 
For one possible embedding of $SU(3)$, two 35-dimensional
representations 
of $SO(8)$ contain three singlets. 
Then the set of 70 scalars in gauged ${\cal N}=8$
supergravity contains six singlets of $SU(3)$. 
It is known \cite{Warner83} 
that $SU(3)$-singlet space with a breaking of the $SO(8)$ gauge group
into a group which contains $SU(3)$
may be written in terms of  the action of $SU(2) \times U(1)$ subgroup
of $SU(8)$ on 70-dimensional representation in the space of self-dual
complex four-forms together with two real parameters.
Instead of taking the
$SU(2)$ group as a subgroup, one can have any $2\times 2$ unitary 
matrix $U(2)$.  
Then this $U(2)$ group element is realized by four real parameters and
moreover there are two additional real parameters.
Recently, a new scalar potential for $U(2) \times U(1)$ subgroup
of $SU(8)$ of gauged ${\cal N}=8$ supergravity has been found in \cite{AW09-1}.
Although $A_1$ tensor of the theory depends on the parameters on
$U(2)$ group, 
after diagonalizing the $A_1$ tensor, the two eigenvalues 
become simple and they can be obtained  from those eigenvalues of 
\cite{AW} by field redefinitions. This new scalar potential
has alternative form that can be read off from the scalar potential in
\cite{Warner83}.
The nontrivial BPS domain-wall solutions for restricted scalar
submanifold from direct extremization of energy-density is also discussed. 

In this paper, we would like to describe the corresponding 
Chern-Simons matter theory in three-dimensions dual to turning on the 
six scalar fields for $SU(3)$-singlet space 
of gauged ${\cal N}=8$ supergravity. The phase structure of the flows 
with two mass parameters was studied in \cite{BHPW} through      
the analysis of gravity dual that corresponds to turn on the four
scalar fields for the $SU(3)$-singlet space.
The family of our flows is driven by three independent mass parameters
from the ${\cal N}=8$ $SO(8)$ theory and is controlled by
two IR fixed points, ${\cal N}=1$ $G_2$-invariant fixed point and 
${\cal N}=2$ $SU(3) \times
U(1)$-invariant fixed point.
We would like to see how the extra mass parameter changes the flows
connecting these three supersymmetric fixed points. This can be seen
from the supergravity dual analysis where the extra two supergravity
fields become the other two fields respectively.   
The generic ${\cal N}=1$ supersymmetric flow with arbitrary mass parameters 
approaches to the ${\cal N}=2$ $SU(3) \times U(1)$
fixed point where the three masses become identical. 
A particular ${\cal N}=1$ supersymmetric $SU(3)$-invariant RG flow
from the ${\cal N}=1$ $G_2$-invariant fixed point to the ${\cal N}=2$
$SU(3)\times U(1)$-invariant fixed point occurs.

In section 2, we review the construction of scalar potential 
given in \cite{AW09-1} for $U(2) \times U(1)$ subgroup
of $SU(8)$ of gauged ${\cal N}=8$ supergravity.  
We focus on the nontrivial supersymmetric critical points   
and describe the BPS domain-wall solutions for restricted scalar
submanifold from direct extremization of energy-density.  
We also consider the supersymmetric flows around three supersymmetric
critical points.

In section 3, we deform the BL theory by generalizing the possible
mass terms preserving the common $SU(3)$ symmetry. We concentrate on 
${\cal N}=1$ and ${\cal N}=2$ supersymmetries. We obtain the deformed 
superpotential with three mass parameters. Based on the supergravity 
dual theory found in section 2, we analyze the features of RG flows in
these parameter space by looking at the role of the extra 
mass parameter. 

In section 4, we summarize what we have obtained in this paper and 
make some comments on the future directions. 

\section{The holographic supersymmetric RG flows  
in four dimensions }

The gauged ${\cal N}=8$ supergravity theory contains self-interaction of a single
massless ${\cal N}=8$ supermultiplet with local $SO(8)$- and  local
$SU(8)$-invariances.   
The 70-real, physical 
scalars of gauged ${\cal N}=8$ supergravity
parametrize the coset space $E_{7(7)}/SU(8)$. 
The most general $SU(3)$-singlet space parametrized by six 
fields(denoted by $\la, \la',\rho, \al, \phi, \varphi$) among 70-fields is
represented by the 56-bein of the fundamental representation of
$E_{7(7)}$.
After exponentiating the $SU(3)$-singlet which is $56 \times 56$
matrix, 
this 56-bein can be decomposed into
two independent 28-beins. From the T-tensor defined by these
28-beins, the new scalar potential \cite{dN82,dN82-1} 
parametrized by the above six fields of the theory
is found in \cite{AW09-1} via long, tedious computations. 
Although the scalar potential looks very complicated at first sight, 
it can be rewritten as the scalar potential \cite{Warner83} 
parametrized by four fields through
simple field redefinitions, according to the observation of \cite{AW09-1}.
This feature makes easier to analyze the critical points of the scalar potential.

As one of the angular variables is equal to the other angular 
variable($\varphi=\phi$),
then the scalar potential with five independent fields has a sum of
the square of superpotential and the squares of derivatives of
superpotential with respect to the fields. 
More explicitly the scalar potential depends on $\la, \sqrt{\la'^2
+\rho^2}, \alpha$ and $\phi$ and the dependence of the extra
field $\rho$ occurs only in the form of $\sqrt{\la'^2 +\rho^2}$.
That is, it depends on four independent quantities. 
From the kinetic terms of
the theory, further constraint on the field, i.e., the condition that 
two fields are equal($\rho=\la'$), gives a simple relation between the scalar
potential and the superpotential because the expression
$\sqrt{\la'^2+\rho^2}$ becomes $\sqrt{2} \la'$ in this case. 
This is required to possess the BPS
bound of the energy-density. Therefore, the extra fields
$(\rho,\varphi)$
we turn on newly in the four-forms \cite{AW09-1} 
should be the same as $(\la',\phi)$  
respectively along the whole RG flows we are considering.
We'll use this property when we discuss the IR behavior of the family of flows
in terms of deformed mass terms in the superpotential in next section.
Finally, the scalar potential reduces to the one \cite{Warner83} with
a simple rescale $\sqrt{2}$ on the field $\la'$.

In summary, the reduced supergravity potential on the most general 
$SU(3)$-invariant sector, by putting the constraints $\varphi=\phi$
and $\rho=\la'$, is then given by \cite{AW09-1}
\bea 
V(\la, \la',\rho; \al, \phi, \varphi)|_{\varphi=\phi,\rho=\la'}= 
g^2 \left[ \frac{16}{3} \left|\frac{\partial
z_3}{\partial \la} \right|^2 + 2 \left| \frac{\partial z_3}{\partial
\la'}\right|^2 - 6  \left|z_3\right|^2 \right],
\label{reducedpotential}
\eea
where the complex superpotential $z_3$ in (\ref{reducedpotential}), 
which is an eigenvalue of $A_1$
tensor of the theory, 
has the following form \cite{AW09-1}
\bea
z_3(\la, \la',\rho; \al, \phi, \varphi) & = & 6e^{i(\alpha + 2\beta)}p^2qr^2t^2
 + 6e^{2i(\alpha + \beta)}pq^2r^2t^2  + p^3(r^4 + e^{4i\beta}t^4)  \nonu \\
& & +
e^{3i\alpha}q^3(r^4 + e^{4i\beta}t^4),
\label{z3}
\eea
and the hyperbolic functions $p,q,r$ and $t$ that depend on $\la$ or $\la'$ and 
trigonometric function $\beta$ are introduced
and they are reduced to as follows after imposing the conditions 
$\varphi=\phi, \rho=\la'$:
\bea 
p  & \equiv &  \cosh (\frac{\la}{2\sqrt{2}}), \qquad q \equiv
\sinh(\frac{\la}{2\sqrt{2}}), \nonu \\
r  & \equiv & 
\cosh(\frac{\sqrt{\la'^2+\rho^2}}{2\sqrt{2}})|_{\rho=\la'}=
\cosh( \frac{\la'}{2}), \qquad
t \equiv \sinh(\frac{\sqrt{\la'^2+\rho^2}}{2\sqrt{2}})|_{\rho=\la'}=
\sinh( \frac{\la'}{2}), 
\nonu \\
\beta & \equiv & \frac{1}{2}
\cos^{-1}( \frac{\la'^2 \cos2\phi + \rho^2
  \cos2\varphi}{\la'^2+\rho^2})|_{\varphi=\phi, \rho=\la'}= \phi. 
\label{pqrt1}
\eea 

There exist six critical points of this scalar potential. Three of
them are supersymmetric while the other three are nonsupersymmetric.
The symmetry group has a common $SU(3)$ group.
Let us present the three supersymmetric ones.

$\bullet$ $SO(8)$ critical point

This occurs at $\lambda = 0 =\lambda'=\rho$, the cosmological constant is
$\Lambda = -6 g^2$(and $W=1$) and 
the ${\cal N}=8$ supersymmetry is preserved.

$\bullet$ $G_2$ critical point

There is a critical point at
$\la=\sqrt{2}\,\sinh^{-1}(\sqrt{\frac{2}{5}(\sqrt{3}-1)})=\frac{\la'}{\sqrt{2}}=
\frac{\rho}{\sqrt{2}}$
as well as $\al=\cos^{-1}(\frac{1}{2}\sqrt{3-\sqrt{3}})=\phi=\varphi$
and the cosmological constant is $\Lambda=-\frac{216 
\sqrt{2}}{25 \sqrt{5}}\cdot 3^{\frac{1}{4}} g^2
$(and $W=\sqrt{\frac{36\sqrt{2} \cdot 3^{\frac{1}{4}}}{25\sqrt{5}}}$).
This  has an unbroken ${\cal N}=1$ supersymmetry.

$\bullet$ $SU(3) \times U(1)$ critical point

Finally, there is a critical point at $\lambda = \sqrt{2} \sinh^{-1}
(\frac{1}{\sqrt{3}}), \, \frac{\lambda'}{\sqrt{2}} = 
\sqrt{2} \sinh^{-1} (\frac{1}{\sqrt{2}})=\frac{\rho}{\sqrt{2}},
\, \alpha=0$ and $\phi=\frac{\pi}{2}=\varphi$
and the cosmological constant is $\Lambda=-\frac{9 \sqrt{3}}{2} g^2$(and
$W=
\frac{3^{\frac{3}{4}}}{2}$).
This critical point has an unbroken ${\cal N}=2$ supersymmetry.

The supersymmetric flow equations together with (\ref{pqrt1}) are
given by \cite{AW09-1}
\bea  
\frac{d \la}{d r} & = &  \frac{8\sqrt{2}}{3}g \partial_{\la} W,
\; \qquad 
\frac{d \la'}{d r}  =   \sqrt{2} g \pa_{\la'} W = \frac{d \rho}{d r}, \nonu \\
\frac{d \alpha}{ d r} & = & 
\frac{\sqrt{2}}{3p^2 q^2} g \pa_{\al} W, \qquad
\frac{d \phi}{d r}    =    \frac{\sqrt{2}}{4r^2 t^2} g \pa_{\phi} W
=\frac{d \varphi}{d r}, \nonu \\ 
\frac{d A}{ d r} & = & - \sqrt{2} g W,
\label{BPS}
\eea
where the real superpotential is given by
\bea
W=|z_3|,
\label{superpotential}
\eea
with (\ref{z3}) and (\ref{pqrt1}).
The scale factor $A(r)$ in the last equation of (\ref{BPS}) 
appears in the four-dimensional metric
$
ds^2= e^{2A(r)} \eta_{\mu \nu} dx^{\mu} dx^{\nu} + dr^2$ with
three-dimensional metric $\eta_{\mu \nu}=(-,+,+)$. The superpotential
$W$ appearing in (\ref{BPS}) is 
the same as the one in \cite{AW} except the factor
$\sqrt{2}$ in front of $\la'$(or $\rho$) for the hyperbolic functions of 
(\ref{pqrt1}). 
Moreover, there exist extra two first-order differential equations 
on $\rho$ and $\varphi$ in (\ref{BPS}).
Let us emphasize that although we have started with six $SU(3)$-singlet fields rather
than four $SU(3)$-singlet fields, it turns out the scalar potential
(\ref{reducedpotential}) 
and the superpotential (\ref{superpotential}) reduce to those in four 
$SU(3)$-singlet fields up to rescale we mentioned before, 
with extra conditions $\varphi=\phi$ and $\rho=\la'$.   
Then the critical points reduce to those for four $SU(3)$-singlet
fields \footnote{The supergravity kink interpolates between an
  asymptotically
$AdS_4$ space in the $UV(r\rightarrow \infty)$ and another in the 
$IR(r\rightarrow -\infty)$ and this can be thought of as an explicit 
construction of a RG flow between a UV fixed point and an 
IR fixed point of the boundary field theory. For $AdS_5$
compactification, see the earlier work \cite{FGPW}.}.

In order to understand the flows around the three supersymmetric fixed points,
${\cal N}=8$ $SO(8)$, ${\cal N}=2$ $SU(3)\times U(1)$ and ${\cal N}=1$ 
$G_2$, one can analyze 
the scaling dimensions of the
operators and the linearization of the flow equations (\ref{BPS}) 
can be done around these
fixed points. From the mass spectrum formula on ${\bf S}^7$, 
the ${\bf
35}_c$ pseudo-scalars  of $SO(8)$ 
can be identified with the second derivative of
the scalar potential evaluated at the $SO(8)$ fixed point and
correspond to the conformal primaries of $\Delta=2$ which consists of
quadratic fermions in the irreducible representations ${\bf 8}_c$ of 
$SO(8)$. 
Here the 
$SO(8)$ coupling constant $g$  is given in
terms of the radius of $AdS_4$ via $g=\frac{1}{\sqrt{2} r_{UV}}$.
The asymptotic behavior of $A(r)$ is given by $A(r) \rightarrow
\frac{r}{r_{UV}} + \mbox{const}$ for $r \rightarrow \infty$.
On the other hand, the ${\bf 35}_v$ scalars of $SO(8)$ correspond to 
the other second derivative of
the scalar potential evaluated at the $SO(8)$ fixed point. The
conformal dimension is $\Delta=1$ which consists of quadratic bosons
in the irreducible representations ${\bf 8}_v$ of $SO(8)$.
The holographic
RG flow equations connecting ${\cal N}=8$ $SO(8)$-invariant fixed point 
to ${\cal N}=2$ $SU(3) \times U(1)$-invariant fixed point were found
in \cite{AP} and
the other holographic
RG flow equations from ${\cal N}=8$ $SO(8)$-invariant fixed point 
to
${\cal N}=1$ $G_2$-invariant
fixed point also were studied in \cite{AI}.

Now the remaining holographic RG flow equations connect between 
${\cal N}=1$ $G_2$-invariant fixed point and 
${\cal N}=2$ $SU(3) \times U(1)$-invariant fixed point.
Compared with the previous RG flow equations in previous paragraph 
where we have dealt with
only two fields, these RG flow equations 
should keep all the four $SU(3)$-singlet fields.  
Recall that the scalar potential
and the superpotential reduce to those in four 
$SU(3)$-singlet fields.
For the flows near the neighborhood of other two supersymmetric fixed
points, ${\cal N}=2$ $SU(3) \times U(1)$-invariant fixed point and
${\cal N}=1$ $G_2$-invariant fixed point,
one can compute the mass eigenvalues at the critical points
\cite{BHPW}.
At the ${\cal N}=2$ $SU(3) \times U(1)$-invariant fixed point, 
the eigenvalues are given by 
$\frac{1 \pm \sqrt{17}}{2}$ and $\frac{3\pm \sqrt{17}}{2}$ which have
two negative values corresponding to irrelevant operators flowing into
the IR fixed point and have two positive values corresponding to
relevant operators or vacuum expectation values driving the flows away
from the IR fixed point. 
At the ${\cal N}=1$ $G_2$-invariant fixed point,
the eigenvalues are 
$1\pm \sqrt{6}$ and $1\pm \frac{1}{\sqrt{6}}$ which have one negative 
value and which have three positive values. 
The two of positive values($1\pm \frac{1}{\sqrt{6}}$),
which are less than $\frac{3}{2}$, 
correspond to non-normalizable modes and should be interpreted as a
perturbation of Lagrangian we are interested in. 
The sum of these eigenvalues is 2.
These represent the fermionic and bosonic
mass terms generating the ${\cal N}=1$ supersymmetric flow from 
the ${\cal N}=1$ $G_2$-invariant fixed point to ${\cal N}=2$ $SU(3)
\times U(1)$-invariant fixed point. They claim that 
the supergravity predicts an anomalous dimension of 
$\pm\frac{1}{\sqrt{6}}$ that drives this ${\cal N}=1$ flow \cite{BHPW}.  

\section{The holographic 
supersymmetric M2-brane flows in three dimensions }

Let us recall that 
the six four-forms consisting of three self-dual and three
anti-self-dual are regarded as six fields in gauged ${\cal N}=8$
supergravity
and they are given explicitly by  \cite{AW09-1,BHPW,HW}
\bea
F_1^{\pm} &=& \varepsilon_{\pm} \left[ \; (\de^{1234}_{IJKL} \pm 
\de^{5678}_{IJKL})+
 (\de^{1256}_{IJKL} \pm \de^{3478}_{IJKL})+(\de^{3456}_{IJKL}
 \pm \de^{1278}_{IJKL}) \; \right],
\nonu \\
       F_2^{\pm} &=& \varepsilon_{\pm} \left[ -(\de^{1357}_{IJKL}
\pm \de^{2468}_{IJKL})+(\de^{2457}_{IJKL} \pm
\de^{1368}_{IJKL})+(\de^{2367}_{IJKL} \pm \de^{1458}_{IJKL}) +
 (\de^{1467}_{IJKL} \pm \de^{2358}_{IJKL}) \right], 
\nonu \\
 F_3^{\pm} &=& \varepsilon_{\pm} \left[ (\de^{2467}_{IJKL}
\mp \de^{1358}_{IJKL}) - (\de^{1367}_{IJKL} \mp
\de^{2458}_{IJKL}) - (\de^{1457}_{IJKL} \mp \de^{2368}_{IJKL}) -
 (\de^{2357}_{IJKL} \mp \de^{1468}_{IJKL}) \right]
\label{four}
\eea
where  $\varepsilon_{+}=1$ for scalars and 
$ \varepsilon_{-} =i$ for pseudoscalars.
The indices $I,J,K,L$  
of these four-forms are the Cartesian coordinates in the
vector representation of $SO(8)$ of gauged ${\cal N}=8$ supergravity theory. 
In particular, the index $7$ and index $8$ in (\ref{four}) which are stabilized by 
the subgroup $SO(6)$ of $SO(8)$ appear simultaneously in
$F_1^{\pm}$(in the indices of $5678,3478$, and $1278$)
while they appear in $F_2^{\pm}$ and $F_3^{\pm}$ independently. 
The $F_2^{\pm}$ are related to the $F_3^{\pm}$ by replacing the index
$7$ with the index $8$ and vice versa up to the signs.
Note that all these four-forms are invariant under the $SU(3)$ subgroup 
of $SO(6) \subset SO(8)$.

The mass deformation of BL theory 
\cite{BL0711,BL06,BL0712}($U(2) \times U(2)$ Chern-Simons
matter theory with level $k=1$ or $k=2$ is equivalent to the two
M2-branes of BL theory) has the following
fermion(32 component Majorana spinors of $SO(1,10)$ subject to a
chiral condition on the M2-brane world volume which leads to 16 real
degrees of freedom) mass terms(See also \cite{GSP,HLL,GRVV})
in the Lagrangian
\bea
{\cal L}_{f.m.} & = & -\frac{i}{2} h_{ab} \bar{\Psi}^a 
\left(  m_1 \Gamma^{78910}+
  m_2\Gamma^{56910}+
  m_3\Gamma^{5678} \right. \nonu \\ 
&- & 
  m_4\Gamma^{46810} 
 +    m_5\Gamma^{4679} +  m_6\Gamma^{4589} +
  m_7\Gamma^{45710} 
\nonu \\
& - & \left.   m_9 \Gamma^{35710} 
 -    m_{10} \Gamma^{3589} -  m_{11} \Gamma^{3679} +
  m_{12} \Gamma^{36810} \right) \Psi^b.
\label{fermionic}
\eea
Intentionally, we put minus signs in the four mass terms corresponding to
the self-dual four-forms that have minus signs in (\ref{four}) above.
We denote only $SO(4)$ gauge index $a$ in the fermion and the spinor
indices for fermions are omitted.
The mass terms in (\ref{fermionic}) consist of fourth order product of the
eleven-dimensional Gamma matrices $\Gamma^{\mu}$ where $\mu =0, 1, 2,
\cdots, 10$ which satisfy the usual
anticommutator relations with the metric.
The spinors are eigenvectors of both $\Gamma^{012}$ and $\Gamma^{345678910}$.
The $SO(8)$ vector indices of (\ref{four}) 
can be seen from (\ref{fermionic}) by subtracting two from those in 
(\ref{fermionic}).
Namely, the indices $78910$ containing $m_1$ in (\ref{fermionic})
correspond to the indices $5678$ of the second term in $F_1^{\pm}$ of 
(\ref{four}). The spinor indices are contracted with those of Gamma matrices.
The presence of $F_3^{\pm}$ in the internal four-forms flux 
corresponding to the supergravity fields $(\rho,\varphi)$ 
in this paper reflects the last line of (\ref{fermionic}) by choosing
the indices $1358, 1367,1457$ and $1468$ from (\ref{four}).

By assuming the linearity in masses, 
the fermionic supersymmetry transformation gets modified by
the following terms which are linear in the mass
\bea
\delta_{mass} \Psi^a  & = & \left(   m_1 \Gamma^{78910}+
  m_2\Gamma^{56910}+
  m_3\Gamma^{5678} 
-  m_4\Gamma^{46810} 
 +  m_5\Gamma^{4679} +  m_6\Gamma^{4589} +
  m_7\Gamma^{45710} 
\right.
\nonu \\
& - &  \left. m_9 \Gamma^{35710} 
 -    m_{10} \Gamma^{3589} -  m_{11} \Gamma^{3679} +
  m_{12} \Gamma^{36810} \right) X_I^a \Gamma^I \epsilon.
\label{mod1}
\eea
The supersymmetry parameter $\epsilon$ in (\ref{mod1}) is 
also eigenvectors of both $\Gamma^{012}$ and $\Gamma^{345678910}$.
Then the linear or quadratic terms in masses to the Lagrangian 
should be determined.  
Let us introduce the bosonic mass terms \cite{GSP,HLL,GRVV} 
\bea
{\cal L}_{b.m.} = -\frac{1}{2} h_{ab} X_I^a  (m^2)_{IJ} X_J^b,
\label{bosonic}
\eea
and we would like to determine $(m^2)_{IJ}$ in (\ref{bosonic}) 
explicitly which will play the role of mass-deformed superpotential
together with (\ref{fermionic}) later. Of course, before the mass
deformation, the superpotential of 
original theory contains the quartic terms in the ${\cal N}=1$
superfield 
having the bosonic field $X_I^a$ as fermionic independent term.  

From the supersymmetry transformation for the bosonic fields $\delta
X_I^a = i \bar{\epsilon} \Gamma_I \Psi^a$ and the supersymmetry
variation for the spinors  (\ref{mod1}), one obtains the quadratic
mass terms in the Lagrangian which contains the bosonic mass terms
(\ref{bosonic}) and the fermionic mass terms (\ref{fermionic}). 
The variation is as follows 
\bea
\delta {\cal L} & = &  i h_{ab} X_I^a  (m^2)_{IJ} \bar{\Psi}^b \Gamma_J \epsilon  
- i h_{ab} \bar{\Psi}^a  \left(   
 m_1 \Gamma^{78910}+
  m_2\Gamma^{56910}+
  m_3\Gamma^{5678} \right. \nonu \\ 
& - &
  m_4\Gamma^{46810} 
 +  m_5\Gamma^{4679} +  m_6\Gamma^{4589} +
  m_7\Gamma^{45710} 
\nonu \\
& - &  \left. m_9 \Gamma^{35710} 
 -    m_{10} \Gamma^{3589} -  m_{11} \Gamma^{3679} +
  m_{12} \Gamma^{36810}
\right)^2
X_I^b \Gamma^I \epsilon. 
\label{van}
\eea
Then the possible supersymmetric mass deformations in (\ref{van}) are characterized
by the bosonic mass terms $(m^2)_{IJ}$, the fourth order product of the
Gamma matrices $\sum_{i} m_i \Gamma_i^{(4)}$ with fermion mass terms
and the supersymmetry parameter $\epsilon$.
In order to vanish the variation (\ref{van}), the relation between the
mass terms of boson and fermion with the supersymmetry parameter 
should be satisfied
\bea
(m^2)_{IJ} \Gamma^J \epsilon  & = &  \left(   
 m_1 \Gamma^{78910}+
  m_2\Gamma^{56910}+
  m_3\Gamma^{5678}  - 
  m_4\Gamma^{46810} 
 +  m_5\Gamma^{4679} +  m_6\Gamma^{4589} +
  m_7\Gamma^{45710} 
\right. \nonu \\
& - &  \left. m_9 \Gamma^{35710} 
 -    m_{10} \Gamma^{3589} -  m_{11} \Gamma^{3679} +
  m_{12} \Gamma^{36810}
\right)^2 \Gamma_I \epsilon.
\label{condition}
\eea
Let us compute the quadratic mass terms in the right hand side of
(\ref{condition}) in order to determine the bosonic mass terms of the
left hand side.
The 38 terms among them are given by
\bea
&& \sum_{i=1}^{7} m_i^2 +  \sum_{i=9}^{12} m_i^2 +  2\left[(-m_1
  m_2-m_4 m_7-m_5 m_6 
- m_{9} m_{12} -m_{10} m_{11})
  \Gamma^{5678} \right. \nonu \\
&& +  (-m_1
  m_3-m_4 m_6-m_5 m_7 
-m_{9} m_{11} - m_{10} m_{12})\Gamma^{56910}
\nonu \\
&& +  (-m_1
  m_4-m_2 m_7+m_3 m_6 
)\Gamma^{4679} + 
(m_1
  m_5+m_2 m_6+m_3 m_7 
)\Gamma^{46810} \nonu \\
&& +  (-m_1
  m_6-m_2 m_5-m_3 m_4 
)\Gamma^{45710}  +  (-m_1
  m_7-m_2 m_4-m_3 m_5 
)\Gamma^{4589}
\nonu \\
&& \left. +
(-m_2
  m_3-m_4 m_5-m_6 m_7 
-m_{9} m_{10} - m_{11} m_{12}) \Gamma^{78910} \right], 
\label{46}
\eea
and the remaining 28 terms are given by
\bea
&& 2\left[(m_1
  m_9+m_2 m_{12}+m_3 m_{11} 
)
  \Gamma^{3589}  +  (m_1
  m_{10}+m_2 m_{11}+m_3 m_{12} 
)\Gamma^{35710} \right.
\nonu \\
&& +  (-m_1
  m_{11}-m_2 m_{10}-m_3 m_9 
)\Gamma^{36810}  + 
(m_1
  m_{12}+m_2 m_{9}+m_3 m_{10} 
)\Gamma^{3679} \nonu \\
&& +  (-m_4
  m_9-m_5 m_{10}-m_6 m_{11} -m_7 m_{12} )
  \Gamma^{345678} \nonu \\
&& +  (-m_4
  m_{10}-m_5 m_9-m_6 m_{12} -m_7 m_{11})\Gamma^{3456910}
\nonu \\
&& +  (-m_4
  m_{11}-m_5 m_{12}-m_6 m_9 -m_7 m_{10})\Gamma^{3478910} \nonu \\
&& \left. + 
(m_4
  m_{12}+m_5 m_{11}+m_6 m_{10} +m_7 m_9)\Gamma^{34} \right]. 
\label{32}
\eea
So far, this holds for any supersymmetry parameter.
From now on we need to classify the possible supersymmetric mass
deformations
by constraining the supersymmetry parameter. 

Let us describe the possible mass deformations based on the number of 
supersymmetry.

$\bullet$ ${\cal N}=1$ supersymmetry 

The $1/8$ BPS condition(the number of supersymmetry is 2)
has the following constraints on the supersymmetry parameter
$\epsilon$ \cite{Ahn0806n1}
\bea
\Gamma^{5678}\epsilon  =  \Gamma^{56910}\epsilon=
\Gamma^{78910}\epsilon = \Gamma^{46810} \epsilon 
=
 -\Gamma^{4679} \epsilon=
-\Gamma^{4589}\epsilon= 
-\Gamma^{45710}\epsilon=
-\epsilon.
\label{n1condition}
\eea 
Then the expression of (\ref{46}) has a simple form when we multiply 
the supersymmetry parameter $\epsilon$ to the right.
Let us apply the Gamma matrices $\Gamma^I$ to (\ref{46}) from the right. 
Then the only nonzero parts are given by
\bea
&& \left[ (m_1+m_2+m_3-m_4 -m_5 -m_6 -m_7 
)^2 + (m_9
  +m_{10}+m_{11} +m_{12})^2 \right] \Gamma^3
\label{3344} 
\\
&&+ \left[ (m_1+m_2+m_3+m_4 +m_5 +m_6 +m_7 
)^2 + (m_9
  +m_{10}+m_{11} +m_{12})^2 \right]\Gamma^4  + \cdots,
\nonu
\eea
where the abbreviated 
$\Gamma^I$($I=5, 6, \cdots, 10$) terms contribute to zero
when the supersymmetry parameter $\epsilon$ is multiplied due to the
constraints (\ref{n1condition}).
Moreover, the first 12 terms of (\ref{32})
are replaced by
\bea
2(m_1+m_2+m_3
)(m_9+ m_{10}+ m_{11} + m_{12}) \Gamma^{3589}, 
\label{3589}
\eea
due to the constraint (\ref{n1condition}) and similarly
the last 16 terms of (\ref{32}) are written as
\bea
-2(m_4+m_5+m_6+m_7)(m_9+ m_{10}+ m_{11} + m_{12}) \Gamma^{345678}. 
\label{345678}
\eea

Then the 33-component of $m^2$ via $   (m^2)_{33} 
\Gamma^3 \epsilon $ can be read off from the coefficient of 
$\Gamma^3$ in (\ref{3344}) and 
 the 44-component of $m^2$ in $  (m^2)_{44} 
\Gamma^4 \epsilon$ can be obtained from the coefficient of 
$\Gamma^4$ in (\ref{3344}).
The off-diagonal 34-component of 
 $m^2$ via  $   (m^2)_{34} 
\Gamma^4 \epsilon $ can be read off from the (\ref{3589}) multiplied by $\Gamma^3$
to the right and 
 the (\ref{345678}) multiplied by $\Gamma^3$
to the right if one identifies the two independent nonzero
components($8$-th and $9$-th components) of
supersymmetry parameter.
Finally, 
the off-diagonal 43-component of 
 $m^2$ in $ (m^2)_{43} 
\Gamma^3 \epsilon $ can be read off from the (\ref{3589}) multiplied by $\Gamma^4$
to the right and 
 the (\ref{345678}) multiplied by $\Gamma^4$
to the right.
In summary, one has the following bosonic mass terms 
\bea
(m^2)_{33} & = &  (m_1+m_2+m_3-m_4 -m_5 -m_6 -m_7 
)^2 + (m_9
  +m_{10}+m_{11} +m_{12})^2,
\nonu \\
(m^2)_{34} & = & -2(m_1+m_2+m_3+m_4+m_5+m_6+m_7
)(m_9+ m_{10}+ m_{11}
+ m_{12}),
\nonu \\
(m^2)_{43} & = & -2(m_1+m_2+m_3-m_4-m_5-m_6-m_7
)(m_9+ m_{10}+ m_{11} + m_{12}),
\nonu \\
(m^2)_{44} & = &
(m_1+m_2+m_3+m_4 +m_5 +m_6 +m_7 
)^2 + (m_9
  +m_{10}+m_{11} +m_{12})^2.
\label{massn1}
\eea
Therefore, for ${\cal N}=1$ supersymmetry, the two mass terms,
$(m^2)_{33}$ and $(m^2)_{44}$,  are
different  due to the mass terms coming from $\sum_{i=4}^7 m_i$
and there are crossed mass terms that will vanish if
$\sum_{i=1}^3 m_i =0$ or $\sum_{i=9}^{12} m_i =0$.
Then the four-fields limit $\sum_{i=9}^{12} m_i =0$ leads to two
independent mass terms in the deformed superpotential as in \cite{BHPW}.
Although we have introduced 11 mass parameters in the fermionic terms
in the Lagrangian, the mass terms (\ref{massn1}) depend on three
independent mass terms, $\sum_{i=1}^3 m_i$ corresponding to the
$F_1^{+}$ four-form, $\sum_{i=4}^7 m_i$ corresponding to the $F_2^{+}$
four-form and 
$\sum_{i=9}^{12} m_i$ corresponding to the $F_3^{+}$ four-form.
We'll come back this issue later.
Thus we have found ${\cal N}=1$ superconformal Chern-Simons matter
theories with global $G_2$ symmetry. We expect that $G_2$-invariant
$U(N) \times U(N)$ Chern-Simons matter theory for $N > 2$ with level
$k=1$ or $k=2$ is dual to the background of $N$ unit of flux. 
We'll see other ${\cal N}=1$ theory with $SU(3)$ global symmetry later.

$\bullet$ ${\cal N}=2$ supersymmetry 

The $1/4$ BPS condition(the number of supersymmetry is 4)
has the following constraints on the supersymmetry parameter
$\epsilon$ \cite{Ahn0806n2}
\bea
\Gamma^{5678}\epsilon & = & \Gamma^{56910}\epsilon=
\Gamma^{78910}\epsilon 
= -\epsilon.
\label{cond}
\eea 
Let us further impose the conditions
\bea
m_1=m_2=m_3
=0.
\label{masscon}
\eea
From the supersymmetry transformation for the bosonic fields $\delta
X_I^a = i \bar{\epsilon} \Gamma_I \Psi^a$ and the supersymmetry
variation for the spinors  (\ref{mod1}) with (\ref{masscon}), one obtains the quadratic
mass terms in the Lagrangian which contains the bosonic mass terms
(\ref{bosonic}) and the fermionic mass terms (\ref{fermionic})
together with (\ref{masscon}). The corresponding expression of
(\ref{van}) becomes 
\bea
\delta {\cal L} & = &  i h_{ab} X_I^a  (m^2)_{IJ} \bar{\Psi}^b \Gamma_J \epsilon  
- i h_{ab} \bar{\Psi}^a  \left(    
  -m_4\Gamma^{46810} 
 +  m_5\Gamma^{4679} +  m_6\Gamma^{4589} +
  m_7\Gamma^{45710} \right.  \nonu \\
& - &  \left. m_9 \Gamma^{35710} 
 -    m_{10} \Gamma^{3589} -  m_{11} \Gamma^{3679} +
  m_{12} \Gamma^{36810}
\right)^2
X_I^b \Gamma_I \epsilon. 
\label{van1}
\eea

In order to vanish this (\ref{van1}), the relation between the bosonic
and fermionic mass terms should be satisfied as before and it is 
given by
\bea
(m^2)_{IJ} \Gamma^J \epsilon  & = &  \left(    
  -m_4\Gamma^{46810} 
 +  m_5\Gamma^{4679} +  m_6\Gamma^{4589} +
  m_7\Gamma^{45710}  \right. \nonu \\
& - &  \left. m_9 \Gamma^{35710} 
 -    m_{10} \Gamma^{3589} -  m_{11} \Gamma^{3679} +
  m_{12} \Gamma^{36810}
\right)^2 \Gamma_I \epsilon.
\label{condition1}
\eea
Let us compute the quadratic mass terms in the right hand side of 
(\ref{condition1}) in order to determine the bosonic mass terms of
left hand side.
The 20 terms which have the structure of Gamma matrices in
(\ref{cond}) 
are given by
\bea
&& \sum_{i=4}^{7} m_i^2 +  \sum_{i=9}^{12} m_i^2 
+  2(-m_4 m_7-m_5 m_6  - m_{9} m_{12} -m_{10} m_{11})
  \Gamma^{5678}  + \cdots, 
\label{20}
\eea
and the remaining 16 terms are
\bea
&& 2(-m_4
  m_9-m_5 m_{10}-m_6 m_{11} -m_7 m_{12} )
  \Gamma^{345678} + \cdots. 
\label{16} 
\eea
Then the expression of (\ref{20}) has a simple form when we multiply 
the supersymmetry parameter $\epsilon$ to the right.
Let us apply the Gamma matrices $\Gamma^I$ to (\ref{20}) from the right.
Then the nonzero parts are given by
\bea
\left[ (m_4 +m_5 +m_6 +m_7)^2 + (m_9
  +m_{10}+m_{11} +m_{12})^2 \right] (\Gamma^3+\Gamma^4) + \cdots,
\label{34}
\eea
where the abbreviated $\Gamma^I$($I=5, 6, \cdots, 10$) terms contribute to zero
when the supersymmetry parameter $\epsilon$ is added due to the
constraints (\ref{cond}).

Moreover, the 16 terms of (\ref{16})
are replaced by
\bea
-2(m_4+m_5+m_6+m_7)(m_9+ m_{10}+ m_{11} + m_{12}) \Gamma^{345678}, 
\label{new345678}
\eea
due to the constraints (\ref{cond}).
The 33-component of $m^2$ in  $   (m^2)_{33} 
\Gamma^3 \epsilon $ can be read off from the coefficient of 
$\Gamma^3$ in (\ref{34}) and 
 the 44-component of $m^2$ in  $   (m^2)_{44} 
\Gamma^4 \epsilon $ can be obtained from the coefficient of 
$\Gamma^4$ in (\ref{34}).
The off-diagonal 34-component of 
 $m^2$ in  $   (m^2)_{34} 
\Gamma^4 \epsilon $  can be found from the (\ref{new345678}) multiplied by $\Gamma^3$
to the right 
if one identifies the four independent nonzero
components($7, 8, 9$-th and $10$-th components in our Gamma matrices
convention) of
supersymmetry parameter.
The off-diagonal 43-component of 
 $m^2$ in  $   (m^2)_{43} 
\Gamma^3 \epsilon $  
can be obtained from the (\ref{new345678}) multiplied by $\Gamma^4$
to the right.
Finally, one has the following bosonic mass terms 
\bea
(m^2)_{33} & = & (m_4 +m_5 +m_6 +m_7)^2 + (m_9
  +m_{10}+m_{11} +m_{12})^2 =(m^2)_{44},
\nonu \\
(m^2)_{34} & = & -2(m_4+m_5+m_6+m_7)(m_9+ m_{10}+ m_{11}
+ m_{12})=-(m^2)_{43}.
\label{massmass}
\eea
Therefore, for ${\cal N}=2$ supersymmetry, the two mass terms,
$(m^2)_{33}$ and $(m^2)_{44}$,  are
equal and there are no crossed mass terms because of $(m^2)_{34} = -(m^2)_{43}$.
Compared with the results of four-fields in \cite{BHPW}, as we take 
the zero limit of $\sum_{i=9}^{12} m_i$, the above mass terms become identical.
The mass terms (\ref{massmass}) depend on two
independent mass terms, $\sum_{i=4}^7 m_i$ corresponding to the $F_2^{+}$
four-form and 
$\sum_{i=9}^{12} m_i$ corresponding to the $F_3^{+}$ four-form.
The presence of $\sum_{i=9}^{12} m_i$ makes the mass terms larger.
Thus we have found ${\cal N}=2$ superconformal Chern-Simons matter
theories with global $SU(3) \times U(1)$ symmetry. Note that 
the ${\cal N}=2$ supersymmetry is encoded in $U(1)$ factor which is R-charge.
It would be interesting to find out $SU(3) \times U(1)$-invariant
$U(N) \times U(N)$ Chern-Simons matter theory for $N > 2$ with level
$k=1$ or $k=2$. 

Let us describe the deformed superpotential by collecting the mass
terms we have found in (\ref{massn1}).
By redefining the masses corresponding to $F_1^{+}, F_2^{+}$ and $F_3^{+}$  
as 
\bea
m_1+m_2+m_3 \equiv M_1, \quad m_4 + m_5 +m_6 +m_7 \equiv M_2, \quad
m_9 +m_{10}+m_{11}+m_{12} \equiv \widehat{m}_3
\label{massred}
\eea
respectively and by introducing 
\bea
M_2 - M_1 \equiv \widehat{m}_1, \qquad 
M_2+M_1 \equiv \widehat{m}_2, 
\label{M1M2}
\eea
one can write down the mass-deformed superpotential in ${\cal N}=1$
superfield notation where the fermionic independent terms are  the bosonic
fields in $\Phi_7 = X_9 + \cdots$ and $\Phi_8=X_{10} +
\cdots $ by reading off the
mass terms (\ref{massn1}) as follows:
\bea
\Delta W = \frac{1}{2} \widehat{m}_7 \Tr \Phi_7^2 +\frac{1}{2} \widehat{m}_8 \Tr \Phi_8^2
+\frac{1}{2} \widehat{m}_{78} \Tr \Phi_7 \Phi_8.
\label{desuper}
\eea
In the action, we should have $\int d^3 x d^2 \theta W \rightarrow 
\int d^3 x d^2 \theta (W + \Delta W)$. 
The ${\cal N}=1$ superpotential $W$ before the deformation 
has quartic terms in $\Phi_I^a$ and comes from the D-term and F-term
of the ${\cal N}=2$ action \cite{BKKS,BHPW}. The supergravity fields 
$(\rho,\varphi)$ we add in section 2 correspond to the dual of 
the operator $\Tr \Phi_7 \Phi_8$. 
Although we have mass terms for the indices $3$ and $4$ in (\ref{massn1})
and (\ref{massmass}),
we present the indices $7$ and $8$ instead in (\ref{desuper}). 
As explained in \cite{AW09-1}, it depends on how one chooses the
$SU(3)$ subgroup of $SO(8)$. If one cooses the index of Gamma matrices
differently, one can make the mass terms appearing the indices $7$ and $8$.
Here the redefined mass terms with (\ref{M1M2}) are given by
\bea
\widehat{m}_7 & \equiv & (M_1-M_2)^2 +\widehat{m}_3^2=\widehat{m}_1^2
+\widehat{m}_3^2, \nonu \\ 
\widehat{m}_8 & \equiv & (M_1+M_2)^2 +\widehat{m}_3^2=\widehat{m}_2^2
+\widehat{m}_3^2, \nonu \\
\widehat{m}_{78} & \equiv &  -4 M_1 \widehat{m}_3= 2 
(\widehat{m}_1-\widehat{m}_2)\widehat{m}_3.
\label{7788}
\eea
Of course, the ${\cal N}=2$ case given in (\ref{massmass}) can be
obtained by taking some constraints on these mass parameters.  

Let us find out the phase structure characterized by the deformed
superpotential (\ref{desuper}) and the coefficients (\ref{7788}).
When we give a mass to two of the fields($\hat{m}_1=0$ and $\hat{m}_2=
2\hat{m}_3$ or
$\hat{m_2}=0$ and $\hat{m}_1=2\hat{m}_3$) in (\ref{desuper}), 
then there exists a $G_2$ symmetry. 
This can be easily seen since in order to get the $G_2$-invariant
critical point and flows one should take the square of scalar field($\la,\alpha$)
in the
coefficient of $F_1^{\pm}$ and the square of scalar field($\la',\phi$) 
in the coefficient of
$F_2^{\pm}$ should be equal to each other 
from the supergravity dual. Then one obtains $M_1 =\pm M_2$ from 
(\ref{massred}).   In terms of $\hat{m}_1$ or $\hat{m}_2$, this
implies that $\hat{m}_1=0$ or $\hat{m}_2=0$ from (\ref{M1M2}).
Furthermore, the section $2$ leads to 
the fact that the square of scalar field in the coefficient of
$F_2^{\pm}$ should be equal to 
the square of scalar field($\rho, \varphi$) in the coefficient of
$F_3^{\pm}$ each other.
Then one obtains $\hat{m}_3 =M_2$ or $\hat{m}_3=M_1$ from 
(\ref{massred}). In terms of $\hat{m}_1$ or $\hat{m}_2$,
this becomes either $\hat{m}_1=0$ and $\hat{m}_2=2\hat{m}_3$ or
$\hat{m_2}=0$ and $\hat{m}_1=2\hat{m}_3$ as above.
We present these RG flows in the three dimensional mass parameter
space in Figure 1 where the
theory from $SO(8)$-invariant fixed point located at the origin 
flows a $G_2$-invariant fixed point when one of the masses is
zero($\hat{m}_1=0$ or $\hat{m}_2$=0) 
and the remaining masses satisfy either  $\hat{m}_2=2\hat{m}_3$
or  $\hat{m}_1=2\hat{m}_3$. For the former $G_2$ fixed point, we have 
$\hat{m}_7=\hat{m}_3^2, \hat{m}_8=5\hat{m}_3^2$ and 
$\hat{m}_{78}=-4\hat{m}_3^2$ from (\ref{7788}) while for the latter
$G_2$ fixed point, there are 
$\hat{m}_7=5 \hat{m}_3^2, \hat{m}_8= \hat{m}_3^2$ and 
$\hat{m}_{78}=4\hat{m}_3^2$. These two fixed points are equivalent to
each other.

When we give an equal mass to three
fields($\hat{m}_1=\hat{m}_2=\hat{m}_3$), 
then there exists $SU(3)
\times U(1)$ symmetry. 
How does one see this? From (\ref{massmass}) and (\ref{7788}), 
one can easily see $\hat{m}_1=\hat{m}_2$ which leads to $M_1=0$
and $M_2=\hat{m}_1=\hat{m}_2$. Also there is no cross term $\hat{m}_{78}=0$.
Furthermore, the section $2$ leads to 
the fact that the square of scalar field($\la', \phi$) in the coefficient of
$F_2^{\pm}$ should be equal to 
the square of scalar field($\rho,\varphi$) in the coefficient of
$F_3^{\pm}$ each other. Then one obtains
$\hat{m}_1=\hat{m}_2=\hat{m}_3$ and  this RG flow(along this flow the
${\cal N}=2$ supersymmetry is encoded in the $U(1)=SO(2)$ factor which is
nothing but R-charge) 
in the three dimensional mass parameter
space is drawn in Figure 1 where 
the
theory from $SO(8)$-invariant fixed point located at the origin 
flows the $SU(3) \times U(1)$-invariant fixed point when three masses
are equal. For $SU(3) \times U(1)$ fixed point, 
we have $\hat{m}_7=2 \hat{m}_3^2=\hat{m}_8$ and $\hat{m}_{78}=0$
consistent with (\ref{massmass}).

From the analysis of previous section, there exists a special ${\cal
N}=1$ supersymmetric RG flow
from the ${\cal N}=1$ $G_2$-invariant fixed point to the ${\cal N}=2$
$SU(3)\times U(1)$-invariant fixed point preserving $SU(3)$ along the
whole flow. 
This flow is triggered by one of the mass 
term($\hat{m}_1$ or $\hat{m}_2$). Starting from the $G_2$-invariant
fixed point with $\hat{m}_2=2\hat{m}_3$ and $\hat{m}_1=0$, one
increases both $\hat{m}_1$ and $\hat{m_3}$ with fixed $\hat{m}_2$
until $\hat{m}_1=\hat{m}_2=\hat{m}_3$. Namely, along the flow there
exists
$\hat{m}_1=2\hat{m}_3 + \mbox{const}$.
Similarly 
starting from the $G_2$-invariant
fixed point with $\hat{m}_1=2\hat{m}_3$ and $\hat{m}_2=0$,
one increases both $\hat{m}_2$ and $\hat{m_3}$ with fixed $\hat{m}_1$
until $\hat{m}_1=\hat{m}_2=\hat{m}_3$.
Along the flow there
exists
$\hat{m}_2=2\hat{m}_3 + \mbox{const}$.
We present these RG flows in the three dimensional mass parameter
space in Figure 1 where 
the
theory from $G_2$-invariant fixed point  
flows the $SU(3) \times U(1)$-invariant fixed point when the masses
of $\hat{m}_1$ or $\hat{m}_2$ are increasing.

When we give an
unequal mass to three fields($\hat{m}_1 \neq \hat{m}_2 \neq
\hat{m}_3$), 
then there is a $SU(3)$ symmetry. 
This is also seen from the supergravity dual analysis
in the sense that for generic values of six fields there is only 
the least $SU(3)$ symmetry.
From the supergravity it is evident that if one has $G_2$ flow with
$\hat{m}_1 \neq 0$ and if one turns on a small value of $\hat{m}_2$,
then the flow is deflected to the $SU(3) \times U(1)$-invariant fixed
point and so $\hat{m}_2$ and $\hat{m}_3$ grow until 
$\hat{m}_1=\hat{m}_2=\hat{m}_3$.  
Similar feature for $G_2$ flow with
$\hat{m}_2 \neq 0$ occurs 
and if one turns on a small value of $\hat{m}_1$, 
the flow is deflected to the $SU(3) \times U(1)$-invariant fixed
point and so $\hat{m}_1$ and $\hat{m}_3$ grow until they reach 
the $SU(3) \times U(1)$-invariant fixed point.
In Figure 1, except the two $G_2$ flow($\hat{m}_1=2\hat{m}_3$ and
$\hat{m}_2=2\hat{m}_3$) 
and the $SU(3) \times U(1)$
flow($\hat{m}_1=\hat{m}_2=\hat{m}_3$) 
starting from the $SO(8)$-invariant fixed point, the generic ${\cal
N}=1$ supersymmetric flows
starting from the $SO(8)$-invariant fixed point or the $G_2$-invariant
fixed point preserve the $SU(3)$ symmetry.  

What happen for the flows with three nonzero unequal 
masses($\hat{m}_1\neq\hat{m}_2\neq\hat{m}_3$)?
Before the mass terms are added, the original superpotential $W$ has 
terms not having $(\Phi_7, \Phi_8)$, terms in linear in $\Phi_7$,
terms in linear in $\Phi_8$ and terms that depend on $\Phi_7$ and
$\Phi_8$. Moreover, the deformed superpotential $\Delta W$ is given by 
(\ref{desuper}).
When we integrate out $(\Phi_7,\Phi_8)$ in $W+\Delta W$ at low energy
scale, 
we obtain
the quartic terms coming from the original $W$ which do not contain
mass parameters and three kinds of sextic terms with three independent
mass parameters which depend on $\hat{m}_7, \hat{m}_8$ and
$\hat{m}_{78}$ by solving the equations of motion for $\Phi_7$ and 
$\Phi_8$ in $W+\Delta W$. Eventhough it is not clear in field theory
that this should flow to the superconformal field theory in the IR but
the gravity dual shows that 
it will flow to the ${\cal N}=2$ 
$SU(3) \times U(1)$-invariant fixed point.
This implies that in the IR the two independent mass parameters are
equal and the other mass parameter vanishes in the resulting
superpotential $\widehat{W}$ as we take 
$\hat{m}_7=\hat{m}_8$ and $\hat{m}_{78}=0$ consistent with (\ref{massmass}). 
Since the ${\cal N}=1$ theory has no R-charge, it is difficult to find
out the phase structure from the field theory alone. However, 
the AdS/CFT provides that it can be used to study the strongly coupled
field theory.  

\begin{figure}[ht]
   \epsfxsize=3.5in 
\centerline{\epsffile{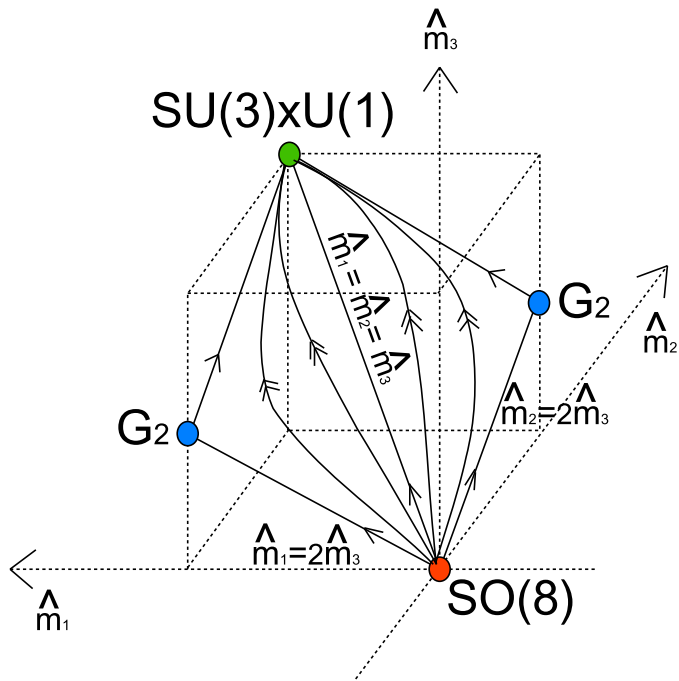}}
   \caption[FIG. \arabic{figure}.]{ 
\sl The RG flows starting from $SO(8)$-invariant fixed point in three
mass parameter spaces. The
theory flows a $G_2$-invariant fixed point when one of the masses is
zero and the remaining masses are nonzero(i.e.,
$\widehat{m}_1=2\widehat{m}_3$ and $\widehat{m}_2=0$ or 
$\widehat{m}_2=2\widehat{m}_3$ and $\widehat{m}_1=0$). If the masses
are equal, then the theory flows to a $SU(3) \times U(1)$-invariant
fixed point($\widehat{m}_1=\widehat{m}_2=\widehat{m}_3$). There are
flows from the two
$G_2$-invariant fixed points to $SU(3)\times U(1)$-invariant fixed
point directly. When the masses are nonzero but not necessarily equal,
the theory flows to the $SU(3) \times U(1)$-invariant fixed
point. Except the flow for equal masses, all the flows are ${\cal
N}=1$ supersymmetric in which there is a $G_2$ symmetry along the flow
ending at the $G_2$-invariant fixed point and otherwise there is a $SU(3)$ 
symmetry.}
\end{figure}

\section{
Conclusions and outlook }

A family of holographic ${\cal N}=1$ supersymmetric RG flows on
M2-branes is studied. 
This family of flows is driven by three independent mass parameters
from maximally supersymmetric ${\cal N}=8$ $SO(8)$ theory and is controlled by
two IR fixed points, $G_2$-invariant fixed point and $SU(3) \times
U(1)$-invariant fixed point. 
The generic flow with different mass parameters is
${\cal N}=1$ supersymmetric and reaches to the $SU(3) \times U(1)$-invariant
fixed point where the three masses become identical and the
supersymmetry is enhanced to ${\cal N}=2$. 
There exists also a special ${\cal N}=1$ supersymmetric RG flow
from the ${\cal N}=1$ $G_2$-invariant fixed point to the ${\cal N}=2$
$SU(3)\times U(1)$-invariant fixed point preserving $SU(3)$ along the
whole flow. 
All these flows are summarized by the Figure 1.
As the title of this paper stands for, we have studied  the
3-dimensional boundary Chern-Simons matter theory explicitly corresponding to the
4-dimensional bulk theory \cite{AW09-1} where there exist six scalar
fields.
In \cite{BHPW}, the phase structure of the flows with two mass
parameters
corresponding to turing on four scalar fields is found.
The extra mass parameter in this paper, compared to \cite{BHPW},
plays an important role.    
When this, corresponding to $\hat{m}_3$ in the Figure 1, 
vanishes, then we reproduce the RG flows obtained in \cite{BHPW}.

It is an open problem to uplift the four-dimensional supergravity
to eleven-dimensional theory. According to \cite{dWNW}, one can construct the
eleven-dimensional metric from the solutions to
four-dimensional gauged ${\cal N}=8$ supergravity.
The nontrivial task is to find out the right expression for the
internal four-form flux which will be present for the 
the most general $SU(3)$-singlet space of
gauged ${\cal N}=8$ supergravity.
So far we have considered the BPS equations (\ref{BPS}) for
$\varphi=\phi, \rho=\la'$
in which the kinetic terms are very simple and 
the scalar potential can be written in terms of a superpotential.
It is natural to ask whether there exist any BPS equations for general vacuum
expectation values or not.  
Besides the two supersymmetric critical points ${\cal N}=2$ $SU(3)
\times U(1)$-invariant fixed point, ${\cal N}=1$ $G_2$-invariant fixed
point of 
four-dimensional gauged ${\cal N}=8$ supergravity,
there are also three nontrivial nonsupersymmetric 
critical points as well as the trivial ${\cal N}=8$ 
$SO(8)$-invariant fixed point for the scalar
potential:$SO(7)^{+}, SO(7)^{-}$ and $SU(4)^{-}$. 
It is an open problem to discover any flow equations connecting any
two (non)supersymmetric fixed points. For the noncompact gaugings 
\cite{AW01,AW02}, it is straightforward to construct the scalar
potential for the six $SU(3)$-singlet space.   

\vspace{.7cm}

\centerline{\bf Acknowledgments}

I would like to thank 
N. Bobev, K. Hosomichi and K. Woo 
for discussions. 

%


\end{document}